\documentclass[11pt,twoside]{article}

\usepackage{asp2006}
\usepackage{epsf}
\usepackage{lscape}
\usepackage{graphicx}

\begin{document}
\title{Discovery of 18 Jupiter mass RV companion
orbiting 
the brown dwarf candidate 
Cha\,H\boldmath{$\alpha$}\,8}   
\author{Viki Joergens and Andr\'e M\"uller}   
\affil{Max-Planck-Institut f\"ur Astronomie, K\"onigstuhl\,17, 69117 Heidelberg, Germany}    

\begin{abstract} 
We report the discovery of a 16--20\,$M_\mathrm{Jup}$ radial velocity companion around the 
very young ($\sim$3\,Myr) brown dwarf candidate Cha\,H$\alpha$\,8 (M5.75--M6.5).
Based on high-resolution echelle spectra 
of Cha\,H$\alpha$\,8 taken between 2000 and 2007 
with UVES at the VLT, a companion was detected through RV variability with a 
semi-amplitude of 1.6\,km\,s$^{-1}$. A Kepler fit to the data 
yields an orbital period of the companion of 1590~days and an eccentricity of $e$=0.49. 
A companion minimum mass $M_2\sin i$ between 16 and 20\,$M_\mathrm{Jup}$ is derived when
using model-dependent mass estimates for the primary.
The mass ratio $q\equiv M_2/M_1$ might be as small as 0.2 and, with 
a probability of 87\%, it is less than 0.4.
Cha\,H$\alpha$\,8 harbors most certainly the lowest mass companion detected so far
in a close ($\sim$ 1\,AU) orbit around a brown dwarf or very low-mass star. 
From the uncertainty in the orbit solution, it 
cannot completely be ruled out that the companion has a mass in the planetary 
regime. 
Its discovery is in any case an important step towards RV planet detections around BDs.
Further, 
Cha\,H$\alpha$\,8 is the fourth known spectroscopic brown dwarf or very low-mass binary system with an
RV orbit solution and the second known very young one. 
\end{abstract}



\section{Introduction}

Search for planetary or brown dwarf (BD) companions to BDs
are of primary interest for understanding planet and BD formation.
There exists 
no widely accepted model for the formation of BDs 
(e.g. Luhman et al.~2007).
The frequency of BDs in multiple systems is a fundamental parameter
in these models.
However, it is poorly constrained for close separations: 
Most of the current surveys for companions
to BDs are done by direct (adaptive optics or HST) imaging and are not sensitive to close 
binaries ($a\la1$\,AU and $a\la10$\,AU for the field and clusters, 
respectively),
and found preferentially close to equal mass systems 
(e.g. Bouy et al.~2003).
Spectroscopic monitoring for radial velocity (RV) variations provides a means to detect 
close systems.
The detection of the first spectroscopic BD binary in the Pleiades, PPl\,15 
(Basri \& Mart\'\i n 1999), raised hope to find many more of these systems in the 
following years. 
However, the number of confirmed close companions to BDs and very low-mass stars (VLMS,
$M \leq 0.1\,M_{\odot}$) is still small.
To date, there are
three spectroscopic BD binaries known, i.e. for which a spectroscopic
orbital solution has been derived: 
PPl\,15, the very young eclipsing system
2M0535-05 (Stassun, Mathieu \& Valenti 2006), and a binary within the quadruple 
GJ\,569 (Zapatero Osorio et al.~2004; Simon, Bender \& Prato 2006). 
They all have a mass ratio close to unity. In particular, no RV planet
of a BD/VLMS has been found yet.
If BDs can harbor planets at a few AU distance is still unknown.
Among the more than 200 extrasolar planets that have been detected around stars
by the RV technique, 6 orbit stellar M-dwarfs 
showing that planets can form also around 
primaries of substantially lower mass than our Sun.
Observations hint that basic ingredients for planet formation
are present also for BDs (e.g. Apai et al.~2005).
However, the only planet detection around
a BD is a very wide 55\,AU system (2M1207, Chauvin et al.~2005), which
is presumably formed very differently from
the Solar System and RV planets.

RV surveys for planets around such
faint objects, as BD/VLMS are, require monitoring with high spectral dispersion
at 8--10\,m class telescopes. While being expensive in terms of telescope
time, this is, nevertheless, extremely important for our understanding
of planet and BD formation.
We report here on the recent discovery of a very low-mass
companion orbiting the BD candidate Cha\,H$\alpha$\,8, which was detected
within the course of an RV survey for (planetary and BD) companions to very young 
BD/VLMS in the Chamaeleon\,I star forming region 
(Joergens \& Guenther 2001; Joergens 2006; Joergens \& M\"uller 2007).


\section{The RV companion of Cha\,H$\alpha$\,8} 

\begin{figure*}[h]
\begin{center}
\includegraphics[width=0.7\linewidth,clip=true,trim=16 16 24 30]{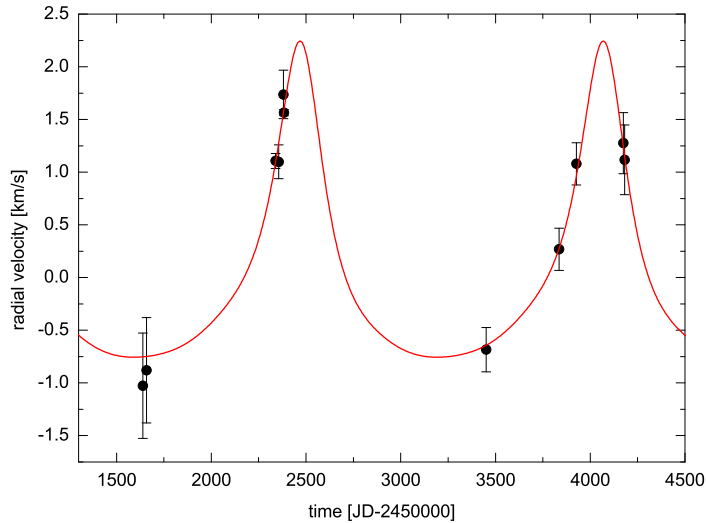} 
\caption{
\label{fig:orbit}
RV measurements of Cha\,H$\alpha$\,8 between 2000 and 2007 based on UVES/VLT spectra.
Overplotted is the best-fit Keplerian orbit, 
which has an RV semi-amplitude of 
1.6\,km\,s$^{-1}$, a period of 4.4\,years and an eccentricity of $e$=0.49.
From Joergens \& M\"uller (2007).}
\end{center}
\end{figure*}

Cha\,H$\alpha$\,8 has been monitored
spectroscopically between 2000 and 2007 with
the Echelle Spectrograph UVES at the VLT 8.2\,m telescope
at a spectral resolution $\lambda$/$\Delta \lambda$ of 40\,000 in the 
red optical wavelength regime. 
RVs were measured based on a cross-correlation technique
employing telluric lines for the wavelength calibration with an accuracy 
of the relative RVs between 30 and 500\,m/s.
As shown in Fig.\,\ref{fig:orbit}, 
the RVs of Cha\,H$\alpha$\,8 are significantly variable on timescales of years
revealing the presence of an RV companion
(Joergens 2006; Joergens \& M\"uller 2007). 
The data allowed us to derive an orbit solution with a reduced $\chi^2$ of 0.42.
This best-fit Kepler orbit has
a mass function of 
4.6 $\times 10^{-4}\,M_{\odot}$, an orbital period of 1590\,d (4.4\,yr), 
an eccentricity of 0.49, and
an RV semi-amplitude of 1.6\,km\,s$^{-1}$. The semi-major axis is of the order of
1\,AU. 
It is noted, that the RV variations cannot be explained by rotational modulation due to
activity  (Joergens \& M\"uller 2007) since the rotation period of Cha\,H$\alpha$\,8 is of the order 
of a few days (Joergens \& Guenther 2001; Joergens et al.~2003).

The mass $M_2 \sin i$ of the companion cannot be determined directly
from a single-lined RV orbit but depends on the primary mass.
Unfortunately, in the case of Cha\,H$\alpha$\,8, the primary mass is not very precisely
determined (as common in this mass and age regime). Using
two available estimates for the primary mass,
0.07\,$M_{\odot}$ (Comer\'on, Neuh\"auser \& Kaas 2000) and 0.10\,$M_{\odot}$ (Luhman 2007),  
$M_2 \sin i$ is inferred to 15.6 and 19.5\,$M_\mathrm{Jup}$, respectively.
This does not take into account errors introduced by evolutionary models.
Based on the assumption of randomly oriented orbits in space, 
the mass ratio q$\equiv$M$_2$/M$_1$ of Cha\,H$\alpha$\,8
is with 50\% probability $\la$0.2  
and with 87\% probability $\la$0.4.
Since the RV technique 
has a bias towards high inclinations, these probabilities can be even higher. 

\section{Discussion and Conclusions}

The companion of Cha\,H$\alpha$\,8 
has most certainly a much smaller mass than that of any previously detected close companion
of a BD/VLMS. 
For comparison, all other known spectroscopic BD binaries have mass ratios $>$
0.6 and the lowest mass BD in these systems has 54\,$M_\mathrm{Jup}$ (Stassun et al.~2006). 
The discovery of the RV companion of Cha\,H$\alpha$\,8 
with its RV semi-amplitude of only 1.6\,km\,s$^{-1}$
is an important step towards RV detections of planets around BD/VLMS.
In fact, from the uncertainty of the orbit solution, it 
cannot be excluded that the companion of Cha\,H$\alpha$\,8
has a mass in the planetary regime ($<13\,M_\mathrm{Jup}$). 
Follow-up RV measurements at the next phase of periastron
will clarify this.

With a semi-major axis of about 1\,AU, the companion of Cha\,H$\alpha$\,8 orbits
at a much closer orbital distance than most companions detected around BDs so far.
In particular, its orbit is much closer than that of  
recently detected very low-mass companions of BDs, like that 
of 2M1207 (55\,AU; Chauvin et al.~2005)
and that of CHXR\,73 (210\,AU; Luhman 2006; see Luhman, this volume).

The favored mechanisms for stellar binary formation, fragmentation of collapsing 
cloud cores
or of massive circumstellar disks, seem to produce preferentially equal mass
components, in particular for close separations
(e.g. Bate et al.~2003).   
Thus, they have difficulties to explain the formation of
the small mass ratio system Cha\,H$\alpha$\,8. However,
we know that close \emph{stellar} binaries with small mass ratios do exist as well
(e.g. q=0.2, Prato et al.~2002), and without knowing the exact mechanism
by which they form, it might be also an option for Cha\,H$\alpha$\,8. 
Considering the small mass of the companion of Cha\,H$\alpha$\,8, 
a planet-like formation could also be possible.
Giant planet formation through core accretion might be hampered
for low-mass primaries, like M dwarfs, by long formation time scales 
(Laughlin, Bodenheimer \& Adams 2004; Ida \& Lin 2005), 
though, recent simulations hint that it can be
a faster process than previously anticipated (Alibert et al.~2005).
On the other hand, giant planets around M dwarfs might form by disk instability
(Boss 2006a, 2006b), at least in low-mass star-forming regions,
where there is no photoevaporation of the disk through nearby hot stars (e.g. Cha\,I).
The companion of Cha\,H$\alpha$\,8 could have been formed through disk instability,
either in situ at 1\,AU or, alternatively, at a larger separation and subsequent
inwards migration. 

Cha\,H$\alpha$\,8 is extremely young (3\,Myr) and 
its study allows insight into the formation and early evolution
at and below the substellar limit. Cha\,H$\alpha$\,8 is only the 2nd known very young BD/VLM 
spectroscopic binary (after 2M0535-05, Stassun et al.~2006).
When combined with 
angular distance measurements or eclipse detections,
spectroscopic binaries allow valuable dynamical mass determinations.
The mass is the most important input parameter for evolutionary models, which rely
for $<$0.3\,M$_\mathrm{\odot}$, only on the two masses 
determined for 2M0535-05.
In order to measure absolute masses of both components of Cha\,H$\alpha$\,8, 
it is required to resolve the spectral lines of both components. 
We will try this with CRIRES/VLT at IR wavelength, were the contrast ratio
between primary and secondary is smaller. 
Having an orbital separation of the order of 13\,milli\,arcsec, 
the spatial resolution of current imaging instruments
is not sufficient to directly resolve Cha\,H$\alpha$\,8.
However, it might be possible to detect the astrometric signal caused by 
the companion, e.g. with NACO/VLT.
This would allow measurement
of the inclination of the orbital plane and, therefore, breaking the 
$\sin i$ ambiguity in the companion mass.


\end{document}